\documentclass[twocolumn,preprintnumbers,amsmath,amssymb]{revtex4}
\usepackage{amsmath,amssymb}
\usepackage{graphicx,epsfig}

\begin{document}
\bibliographystyle {plain}

\def\oppropto{\mathop{\propto}} 
\def\opsimeq{\mathop{\simeq}}
\def\opoverderline{\mathop{\overline}}
\def\operarrow{\mathop{\longrightarrow}}
\def\opsim{\mathop{\sim}}

\def\fig#1#2{\includegraphics[height=#1]{#2}}
\def\figx#1#2{\includegraphics[width=#1]{#2}}


\title{ Non equilibrium dynamics of disordered systems : understanding 
the broad continuum of relevant time scales 
via a strong-disorder RG in configuration space.
   } 


 \author{ C\'ecile Monthus and Thomas Garel }
  \affiliation{Institut de Physique Th\'{e}orique,
 CNRS and CEA Saclay, 91191 Gif-sur-Yvette cedex, France}

\begin{abstract}

We show that an appropriate description of
the non-equilibrium dynamics of disordered systems is obtained through
 a strong disorder renormalization procedure in {\it configuration space},
that we define for any master equation with transitions rates 
$W \left( { \cal C} \to { \cal C}' \right)$ between configurations.
 The idea is to eliminate iteratively the configuration with the highest
 exit rate $W_{out} ({ \cal C} )=
 \sum_{{ \cal C}' } W \left( { \cal C} \to { \cal C}' \right)$
to obtain renormalized transition rates between the remaining configurations.
The multiplicative structure of the new generated transition rates suggests
 that, for a very broad class of disordered systems,
 the distribution of renormalized exit barriers defined as $B_{out} ({\cal C} )
\equiv - \ln W_{out}({\cal C} )$ will become 
broader and broader upon iteration, 
so that the strong disorder renormalization procedure
should become asymptotically exact at large time scales.
We have checked numerically this scenario for 
the non-equilibrium dynamics of a directed polymer in a two dimensional random medium.

\end{abstract}

\maketitle

The non-equilibrium dynamics of disordered systems usually displays 
a broad continuum of relevant time scales, that give rise to
a lot of striking properties such as aging, rejuvenation and memory 
that have been studied a lot both experimentally and theoretically
(see \cite{bouchaud} and references therein).
In finite dimensions, these effects can be understood
 via the growth of some coherence length $l(t)$
that separates the smaller lengths $l < l(t)$ which are quasi-equilibrated
from the bigger lengths $l > l(t)$ which are completely
out of equilibrium. The slow nature of the dynamics then reflects the fact 
that equilibration on larger length scales
requires to overcome larger and larger barriers. 
Within the droplet scaling theory
proposed both for spin-glasses \cite{heidelberg,Fis_Hus} 
and for the directed polymer in a
random medium \cite{Fis_Hus_DP}, the non-equilibrium dynamics
 is activated with barriers scaling as power law $B(l) \sim l^{\psi}$
of the length scale $l$.
The typical time associated 
to scale $l$ then grows as an exponential
$\ln t_{typ}(l) \sim B(l) \sim l^{\psi}$, or equivalently, the characteristic
length-scale $l(t)$ associated to time $t$ grows only logarithmically
$l(t) \sim \left( \ln t \right)^{1/\psi}$.
In numerical studies, this logarithmic behavior has
 remained controversial because
the maximal equilibrated length $l_{max} $ measured
 at the end of the simulations is usually rather small,
so that many fits of the data are possible.
For instance, in Monte-Carlo simulations of 2D or 3D random ferromagnets
 \cite{NUMErandomferro} or spin-glasses \cite{NUMEspinglass},
the maximal equilibrated size is usually only of order $l_{max} \sim 10$.
A noteworthy exception is the simulation of an elastic line in a random medium 
where sizes of order $l_{max} \sim 100 $ have been measured \cite{rosso}
with the conclusion that the length $l(t)$ grows logarithmically 
with a barrier exponent $\psi \sim 0.49$ ( whereas power-law fits $l(t) \sim t^{1/z}$ are excluded at large times).
Note that the difficulties met in dynamical Monte Carlo simulations of
disordered systems comes precisely from the presence
 of a continuum of relevant time scales
ranging from the microscopic scale of single moves to the equilibrium
 time of the full system. Then even faster-than-the-clock Monte Carlo 
algorithms  \cite{algoBKL}, where each iteration leads to a movement, 
 become inefficient 
 because they face the 'futility' problem \cite{werner} :
the number of different configurations visited during the simulation 
remains very small with respect to the accepted moves.
The reason is that the system visits over and over again 
the same configurations within a given valley before it is able to
escape towards another valley where it will be trapped even longer!

In this paper, we argue that the appropriate description of 
these dynamics with a broad continuum of relevant time scales
requires some renormalization where the smaller 
time scales are successively integrated out to obtain
the properties of the large-time dynamics. Moreover, 
since we expect that in these systems, disorder dominates
 over thermal fluctuations
at large scales, the most appropriate 
renormalization scheme is a so called 'strong disorder renormalization (RG)
procedure' (see \cite{review} for a review).
This very specific type of RG, which was  
introduced by Ma and Dasgupta \cite{madasgupta} 
and developed by D.S. Fisher \cite{dsf} in the field of quantum spin chains,
has been then successfully applied to
various classical disordered dynamical models,
such as random walks in random media \cite{sinairg},
reaction-diffusion in a random medium \cite{readiffrg}, 
coarsening dynamics of classical spin chains \cite{rfimrg}, 
trap models \cite{traprg}, 
absorbing state phase transitions \cite{contactrg},
zero range processes \cite{zerorangerg}, 
exclusion processes  \cite{exclusionrg}.
In all these cases, the strong disorder RG rules 
have been formulated {\it in real space},
with specific rules depending on the problem.
In this paper, we show that for more complex systems where
 the formulation of strong disorder RG rules
has not been possible in real space, 
it is nevertheless possible to formulate strong disorder RG rules
{ \it in configuration space}. Moreover, 
this formulation in configuration space
 is very general since it can be defined for any master
 equation describing the evolution of the
probability $P_t ({\cal C} ) $ to be in  configuration ${\cal C}$
 at time t
\begin{eqnarray}
\frac{ dP_t \left({\cal C} \right) }{dt}
= \sum_{\cal C '} P_t \left({\cal C}' \right) 
W \left({\cal C}' \to  {\cal C}  \right) 
 -  P_t \left({\cal C} \right) W_{out} \left( {\cal C} \right)
\label{master}
\end{eqnarray}
The notation $ W \left({\cal C}' \to  {\cal C}  \right) $ 
represents the transition rate per unit time from configuration 
${\cal C}'$ to ${\cal C}$, and 
\begin{eqnarray}
W_{out} \left( {\cal C} \right)  \equiv
 \sum_{ {\cal C} '} W \left({\cal C} \to  {\cal C}' \right) 
\label{wcout}
\end{eqnarray}
represents the total exit rate out of configuration ${\cal C}$.
The two important properties of this master equation are the following
(i)  the exit time $\tau$ from configuration  ${\cal C}$
is distributed with the exponential law
$P^{exit}_{\cal C} (\tau) = W_{out} \left( {\cal C} \right) e^{ -  \tau W_{out} \left( {\cal C} \right)} $
(ii) the new configuration ${\cal C}'$
where the jumps jumps at time $\tau$
when it leaves the configuration ${\cal C}$
is chosen with the probability $\pi_{\cal C} \left({\cal C}' \right)=\frac{W \left({\cal C}  \to {\cal C}' \right)}{W_{out} \left( {\cal C} \right)}$.

For dynamical models, the aim of any renormalization procedure
is to integrate over 'fast ' processes to obtain effective properties 
of 'slow' processes. The general idea of 'strong renormalization' consists 
in eliminating iteratively the 'fastest' process.
For the master equation of Eq. \ref{master},
we thus define the strong disorder renormalization in configuration space 
by the iterative elimination of  the configuration with the highest
exit rate (Eq. \ref{wcout}).
Let us call this configuration ${\cal C}^*$, and its exit rate $W^*_{out}$
\begin{eqnarray}
W^*_{out} = W_{out} \left( {\cal C}^* \right) \equiv 
 {\rm max}_{{\cal C}} \left[  W_{out} \left( {\cal C} \right) \right]
\label{defwmax}
\end{eqnarray}
We now have to compute the 'new' effective transitions rates 
$W^{new}({\cal C} \to {\cal C} ')$
among the remaining configurations in terms of the 'old' transitions rates 
$W^{old}({\cal C} \to {\cal C} ')$
where the decimated configuration ${\cal C}^*$ was still present.
The only changes occur for the configurations called here 
$({\cal C}_1,{\cal C}_2,...,{\cal C}_n)$
that were related via positive rates $W^{old}({\cal C}^* \to {\cal C}_i )>0$
 and $W^{old}({\cal C}_i\to {\cal C}^*)>0$
to the decimated configuration ${\cal C}^*$
(here we will assume, for the simplicity of the discussion, 
and because it is usually the case in
statistical physics models, that if a transition has a strictly positive rate, 
the reverse transition has also
a strictly positive rate; but of course the renormalization rules can be simply
 extended to other cases).
The $2 n$ rates $W^{old}({\cal C}^* \to {\cal C}_i )$ and 
$W^{old}({\cal C}_i \to {\cal C}^*)$
with $i=1,..,n$ have to be eliminated, after taking into account
their effects on transitions between pairs of neighbors of ${\cal C}^*$.
For each neighbor configuration ${\cal C}_i$ with
$i \in (1,..,n)$, the renormalized rate to go to
the configuration ${\cal C}_j$ with $j \in (1,..,n)$ and $j \neq i$ reads
\begin{eqnarray} \nonumber 
&& W^{new}({\cal C}_i \to {\cal C}_j ) =W^{old}({\cal C}_i \to {\cal C}_j ) \\
&& + W^{old}({\cal C}_i \to {\cal C}^* ) \times 
\frac{W^{old}({\cal C}^* \to {\cal C}_j )}
{W^{*}_{out} }
\label{wijnew}
\end{eqnarray}
The first term represents the 'old' transition rate (possibly zero),
whereas the second term represents the transition 
via the decimated configuration ${\cal C}^*$ :
the factor $W^{old}({\cal C}_i \to {\cal C}^* ) $ takes into account 
the transition rate to ${\cal C}^*$,
whereas the second factor represents the probability
 to make a transition towards ${\cal C}_j$
when in ${\cal C}^*$. 
Note that the rule of Eq. \ref{wijnew} 
has been recently proposed in \cite{vulpiani}
to eliminate 'fast states'  from various dynamical problems 
with two very separated time scales.
The physical interpretation of this rule is as follows :
the time spent in the decimated configuration ${\cal C}^*$ is neglected
with respects to the other time scales remaining in the system. 
The validity of this approximation within the present renormalization procedure
will be discussed in detail below. 
To finish the decimation of ${\cal C}^*$, we have now
 to update the exit rates out of the neighboring 
configurations ${\cal C}_i$ 
\begin{eqnarray}
W^{new}_{out}({\cal C}_i)   = 
\sum_{\cal C} W^{new}({\cal C}_i \to {\cal C} )
\label{wioutnewactualisation}
\end{eqnarray}
Since the only changes come from 
the rate towards ${\cal C}^*$ that has disappeared and 
from the rates towards $j \in (1,..,n)$ with $j \neq i$ 
that have changed according to Eq. \ref{wijnew}, one obtains
\begin{eqnarray}
 \nonumber W^{new}_{out}({\cal C}_i) = W^{old}_{out}({\cal C}_i)  
  - W^{old}({\cal C}_i \to {\cal C}^* ) 
\frac{ W^{old}({\cal C}^* \to {\cal C}_i )}{W^{*}_{out}}
\end{eqnarray}
The physical meaning of this rule is the following.
The exit rate out of the configuration ${\cal C}_i$ decays because 
the previous transition towards ${\cal C}^*$ could lead to an immediate return
towards ${\cal C}_i$ with probability 
$\frac{ W^{old}({\cal C}^* \to {\cal C}_i )}
{W^{*}_{out}} $. After the decimation of the configuration ${\cal C}^*$,
this process is not considered as an 'exit' process anymore, but as a
residence process in the configuration ${\cal C}_i$.
This point is very important to understand the
 meaning of the RG procedure :
the remaining configurations at a given stage are 
'formally' microscopic configurations
 of the initial master equation,
but each of these remaining microscopic configuration
 actually represents some 'valley' in configuration space
that takes into account all the previously decimated configurations.
Note that in practice, the renormalized rates $W({\cal C} \to {\cal C}' )$
can rapidly become very small as a consequence of the multiplicative structure
of the renormalization rule of Eq \ref{wijnew}. So the appropriate variables 
are the logarithms of the transition rates, called 'barriers' from now on.
 The barrier $B ({\cal C} \to {\cal C}' )$ from ${\cal C}$  to ${\cal C}' $ 
is defined by
$B ({\cal C} \to {\cal C}' )\equiv - \ln W({\cal C} \to {\cal C}' )$ 
and similarly the exit barrier out of configuration ${\cal C}$ is defined by 
\begin{eqnarray}
B_{out} ({\cal C} )\equiv - \ln W_{out}({\cal C} )
\label{defbout}
\end{eqnarray}
As mentioned above, the approximation made in the
 renormalization rule of Eq. \ref{wijnew}
consists in neglecting the time spent
 in the decimated configuration ${\cal C}^*$ 
with respects to the other time scales remaining in the system. 
In our present framework, this means that the maximal exit
 rate chosen in Eq \ref{defwmax}
should be well separated from the exit rates of the neighboring configurations ${\cal C}_i$.
The crucial idea of 'infinite disorder fixed point' \cite{dsf,review}
is that even if this approximation is not perfect during 
the first steps of the renormalization,
this approximation will become better and better at large RG scale
 if the probability distribution of the remaining exit rates
becomes broader and broader upon iteration. 
More precisely,  if the renormalization scale $\Gamma$ is defined 
as the exit barrier of the last eliminated configuration $\Gamma= B_{out} ({\cal C^*} )$,
one expects that the probability distribution of the remaining 
exit barriers $B_{out} \geq \Gamma$ will 
converge towards some scaling form
\begin{eqnarray}
P_{\Gamma} ( B_{out}-\Gamma   ) \opsimeq_{ \Gamma \to \infty} 
  \frac{1}{\sigma(\Gamma) } {\tilde P} \left( \frac{B_{out} - \Gamma}{\sigma(\Gamma) } \right)
\label{pgammabout}
\end{eqnarray}
where ${\tilde P} $ is the fixed point probability distribution,
 and where $\sigma(\Gamma)$
 is the appropriate scaling factor.
The notion of 'infinite disorder fixed point' means that 
the width $\sigma(\Gamma)$ grows to infinity with the RG scale $\Gamma$.
For instance, in previously known cases of infinite disorder fixed points 
where calculations can be done explicitly \cite{review},
  the scale $\sigma(\Gamma)$ grows linearly  $\sigma(\Gamma) \sim \Gamma$,
 and the fixed point distribution is an exponential  ${\tilde P}(x)=e^{-x}$.
Whenever the flow is towards an 'infinite disorder fixed point', the strong disorder renormalization
procedure becomes asymptotically exact at large RG scales.
For our present problem, the convergence 
towards an 'infinite disorder fixed point' will depend on the initial condition of the transition rates,
i.e. on the model ( and on the temperature if there are phase transitions).
 However, the form of the renormalization rule of Eq \ref{wijnew}
 is sufficiently similar to the usual Ma-Dasgupta rules \cite{review} to think that 
 the convergence towards some infinite disorder fixed point should be realized in
 a very broad class of disordered systems in their glassy phase.
In practice, it can be checked numerically for each model of interest.

\begin{figure}[!ht]
\begin{center}
\includegraphics[width=3.2in]{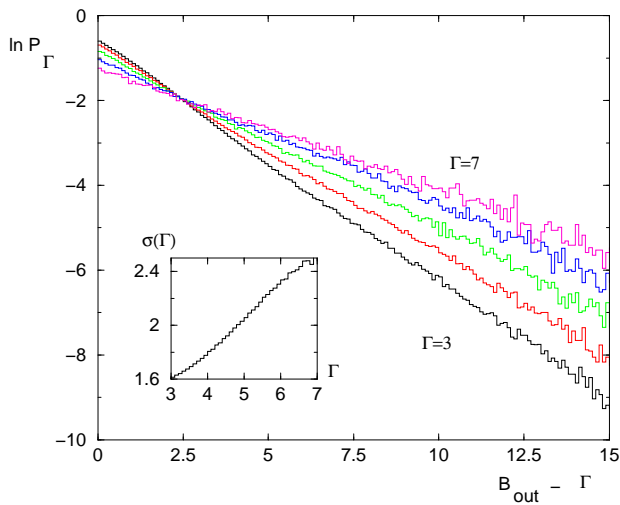}
\end{center}
\caption{
\label{figflow} (Color online)
Flow of the probability distribution $P_{\Gamma}(B_{out}-\Gamma)$
of the renormalized exit barriers (see Eq. \ref{pgammabout})
as the RG scale grows $\Gamma=3,4,5,6,7$.
Inset : growth of the width $\sigma(\Gamma)$ with the RG scale $\Gamma$.
(Data obtained for a directed polymer of length $L=9$
 in a two-dimensional random medium with $2^L=512$ configurations,
 the statistics is over $n_s=15.10^4$ disordered samples).
  }
\end{figure}

As an example of application, we consider 
the directed polymer in a two-dimensional random medium,
a model first introduced to describe interfaces in random ferromagnets
 \cite{Hus_Hen} (see \cite{Hal_Zha} for a review).  The
statics is well described by the Fisher-Huse droplet theory
 \cite{Fis_Hus_DP}
as checked by detailed numerical studies \cite{Fis_Hus_DP,DPexcita}.
We use the following discrete formulation for
 a polymer of length $L$ attached at the origin :
the $2^L$ configurations $(h_1,..,h_L)$ correspond to the random walks
$h_x- h_{x-1} = \pm 1 $ starting at $h_0=0$.
The energy  of a configuration is $E  = \sum_{x=1}^L \epsilon (x,h_x)$
 where the site random energies $\epsilon (x,h)$ are Gaussian
$\rho( \epsilon)= e^{-\epsilon^2/2}/\sqrt{2 \pi}$.
We consider the usual Metropolis dynamics at temperature $T=0.5$ defined by
 the transition rates 
\begin{eqnarray}
W \left( {\cal C} \to {\cal C}'  \right)
= \delta_{<{\cal C}, {\cal C}' >} 
\  {\rm min} \left(1, e^{-  (E({\cal C}' )-E({\cal C} ))/T } \right)
\nonumber 
\label{metropolis}
\end{eqnarray}
where the factor $\delta_{<{\cal C}, {\cal C}' >}$ 
means that the two configurations are related via the move of a single monomer $h_x \to h_x \pm 2$.
So initially each configuration 
has at most $L$ neighbors corresponding to single moves.
However during the renormalization, 
many new transition rates will be generated via Eq. \ref{wijnew},
as in real-space strong disorder RG studies of quantum models 
in dimension $d>1$ \cite{motrunich,lin}.
Here, to validate our approach, we have decided to follow exactly
 the full RG flow, without disregarding any new transition rate. 
 As a consequence, we have been able to study numerically only moderate
 lengths $ L \leq 9$ ( $2^L \leq 512$ configurations )
with a sufficient statistics of $n_s \geq 10^5$ disordered samples
(we have data up to $L=11$ with $n_s=500$ samples, 
but histograms are too noisy).

\begin{figure}[!ht]
\begin{center}
\includegraphics[width=3.2in]{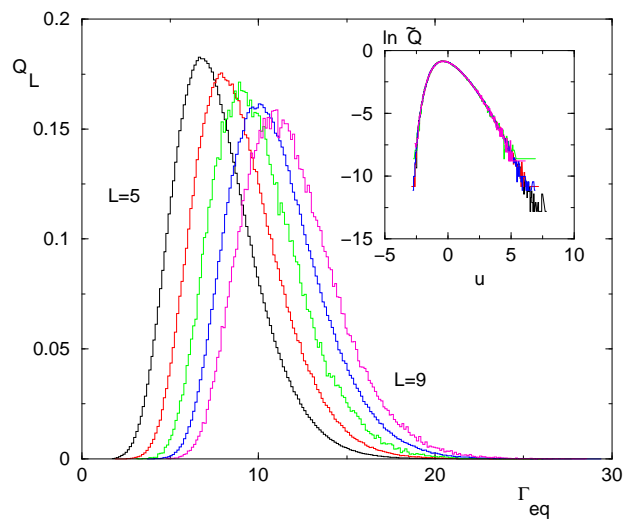}
\end{center}
\caption{
\label{figeq} (Color online)
Statistics of the equilibrium time over the samples
for a directed polymer in a two-dimensional random medium   :
probability distribution $Q_{L}(\Gamma_{eq}=\ln t_{eq})$
for length $L=5,6,7,8,9$ (see Eq. \ref{qlteq}).
Inset : rescaled distribution ${\tilde Q}(u)$
 in log representation to see the tails. }
\end{figure}

We find  (see Fig 1) that the probability distribution
of renormalized exit barrier of Eq. \ref{pgammabout}
 flows towards an ``infinite disorder''
fixed point, with a width growing as $\sigma(\Gamma) \sim \Gamma$
and a rescaled probability which is extremely close to the exponential
${\tilde P} (x) \sim e^{-x}$. Note that this type of renormalized distribution
 seems extremely robust within strong disorder RG
since they hold for exactly in soluble models in $d=1$  \cite{review}
and have been also found numerically in quantum models 
in dimension $d>1$ \cite{motrunich}.
In a finite sample, the typical equilibrium time $t_{eq}$ can be obtained
 from the last decimated barrier $\Gamma_{eq}$
leading to a single surviving configuration via $\Gamma_{eq}= \ln t_{eq}$.
Its probability distribution  $Q_{L}(\Gamma_{eq}=\ln t_{eq})$
over the disordered samples of size $L$ is shown on Fig. 2 
for various $L$.
The convergence towards a fixed rescaled distribution 
\begin{eqnarray}
Q_{L}(\Gamma_{eq}) \sim  
  \frac{1}{\Delta(L) } {\tilde Q} 
\left( u \equiv \frac{\Gamma_{eq} - \overline{\Gamma_{eq}(L)} }{\Delta(L) }
 \right)
\label{qlteq}
\end{eqnarray}
 is rapid
(see inset of Fig. 2) 
but the sizes studied are not sufficient to obtain,
 via the average $\overline{\Gamma_{eq}(L)} \sim L^{\psi}$ or
the width $\Delta(L)\sim L^{\psi}$,
a reliable measure of the asymptotic barrier exponent $\psi$, 
whose value has remained controversial
(see \cite{conjecturepsi} for a recent summary).
We hope in the future to propose simplified ways of following 
the important rates of the renormalized flow
to reach bigger system sizes \cite{future}. 
We also intend to study other properties of the RG flow, 
in particular the structure of the evolving set of remaining configurations
 that label the metastable valleys above a given life-time,
 as well as aging properties. 

In conclusion, we have proposed to describe
the non-equilibrium dynamics of disordered systems via
 a strong disorder renormalization procedure in {\it configuration space},
that we have defined for any master equation.
We have argued that for a very broad class of disordered systems,
 the distribution of renormalized exit barriers 
should become broader and broader upon iteration, 
so that the strong disorder renormalization procedure
should become asymptotically exact at large time scales.
We have checked this scenario numerically for 
the directed polymer in a two-dimensional random medium
and we intend to study other disordered models in the future \cite{future}.

\end{document}